
\input harvmac
\def\ud{\half}

\def\ee#1{{\rm e}^{^{\textstyle#1}}}
\def\d{{\rm d}}
\def\e{{\rm e}}

\def\inv{^{\raise.15ex\hbox{${\scriptscriptstyle -}$}\kern-.05em 1}}
\def\foot#1{\footnote{$^*$}{#1}}
\def\ee#1{{\rm e}^{^{\textstyle#1}}}

\def\frac#1#2{{\textstyle{#1\over#2}}}
\def\title#1{\centerline{\titlefont #1}}
\def\preprint{\Title{\vbox{\baselineskip12pt\hbox{SPhT/92-163}}}}
%

%
\preprint{SPhT/92-163}
\title{Large Order Behaviour of 2D Gravity Coupled to $d<1$ Matter}

\centerline{B. Eynard and J. Zinn-Justin$^1$}
\vskip1.3truecm
{\baselineskip14pt\centerline{Service de Physique Th\'eorique$^2$ de Saclay}
\centerline{F-91191 Gif-sur-Yvette Cedex, FRANCE}}
\footnote{}{$^1$zinn@amoco.saclay.cea.fr}
\footnote{}{$^2$Laboratoire de la Direction des Sciences de la Mati\`ere du
Commissariat \`a l'Energie Atomique}
\vskip0.3truecm
We discuss the large order behaviour and Borel summability of the topological
expansion of models of 2D gravity coupled to general $(p,q)$
conformal matter. In a previous work it was proven that at large order
$k$ the string susceptibility had a generic $a^k\Gamma(2k-\ud)$ behaviour.
Moreover the constant $a$, relevant for the problem of Borel summability,
was determined for all one-matrix models. We here obtain a set of equations
for this constant in the general $(p,q)$ model. String equations can be
derived from the construction of two differential operators $P,Q$ satisfying
canonical commutation relations $[P,Q]=1$. We show that the equation for $a$
is determined by the form of the operators $P,Q$ in the spherical or
semiclassical limits. The results for the general one-matrix models are then
easily recovered.  Moreover, since for the $(p,q)$ string models such
$p=(2m+1)q\pm1$ the semiclassical forms of $P,Q$ are explicitly known, the
large order behaviour is completely determined. This class
contains all unitary $(q+1,q)$ models for which the answer is specially
simple. As expected we find that the topological expansion for unitary models
is not Borel summable.
\Date{26/12/92}

\nref\rDS{M. Douglas and S. Shenker, {\it Nucl. Phys.} B335 (1990) 635.}%
\nref\rBK{E. Br\'ezin and V. Kazakov, {\it Phys. Lett.} B236 (1990) 144.}%
\nref\rGM{D. Gross and A. A. Migdal, {\it Phys. Rev. Lett.} 64 (1990) 127;
{\it Nucl. Phys.} B340 (1990) 333.}%
\newsec{Introduction}
We report here new results concerning the large order
behaviour of the perturbation series of models
\refs{\rDS{--}\rGM}\ of 2D gravity coupled to $D<1$ matter.
Our motivation is to gather some information about non-perturbative features
of quantum gravity and string theory studying the asymptotic behaviour of
perturbation series. In particular we want to understand  whether perturbation
theory indeed provides a proper definition of the physical theory of interest,
in more technical terms whether the perturbation series is Borel summable.

We first recall that the coupled differential equations for the partition
function in the formulation  of 2D quantum gravity coupled to arbitrary
$(p,q)$ minimal conformal matter can be derived from canonical commutation
relations $[P,Q]=1$ \ref\rD{M. R. Douglas, {\it Phys. Lett.} B238 (1990) 176.}
where $P,Q$ are two differential operators of degree $p$ and $q$ respectively:
\eqn\eopPQ{ P=\d^p-\ud\sum_{i=1}\{u_i(x),\d^{p-2i}\}, \qquad
Q=\d^q-\ud\sum_{i=1}\{v_i(x),\d^{q-2i}\},}
and $u(x)=u_1(x)/p=v_1(x)/q$ is the specific heat or string susceptibility.
Note that our normalization of $u(x)$ differs by a factor 2 from the most
commonly used in this problem (this normalization corresponds in the
one-matrix case to consider potentials which are not even).  In this way the
double pole of smallest residue of $u(x)$ has residue 1. Since the partition
function $F$ is given by $F''(x)=-u(x)$, $\e^F$ has then simple zeros. \par
When one of the operators is given it can be shown that the other operator can
be taken of the form:
$$P=Q^{p/q}_+$$
where the subscript $+$ means that $P$ is the sum of the terms of non-negative
power in the formal expansion of $Q^{p/q}$ for $\d$ ``large". In \ref\rGGPZ{P.
Ginsparg, M. Goulian, M. R. Plesser, and J. Zinn-Justin, {\it Nucl. Phys.}
B342 (1990) 539.}\ (see also \ref\rJY{A. Jevicki and T. Yoneya, {\it Mod.
Phys. Lett.} A5 (1990) 1615.}), it was shown that the coupled differential
equations  also follow from an action principle.  The basic action for a
critical $(p,q)$ model takes the general form
\eqn\egenac{S=\int\d x\,\left({\rm Res}\,Q^{p/q+1}+xu\right),}
%
where Res denotes the residue (coefficient of $d^{-1}$) of its fractional
powers.\par
In the simple one-matrix case the ``string equation'' for the specific heat
$u(x)$ \refs{\rDS{--}\rGM} reduces to:
\eqn\ede{(l+\half)R_{l}[u]=x\, ,}
where the $R_{l}$'s are the usual KdV potentials \ref\rGDi{I. M. Gel'fand
and L. A. Dikii, {\it Russian Math. Surveys} 30:5 (1975) 77\semi
I. M. Gel'fand and L. A. Dikii, {\it Funct. Anal. Appl.} 10 (1976) 259.}.
Due to the elementary properties of the $R_l$'s, the above equation follows
as the variational derivative with respect to $u$ of the action
\eqn\egdp{S=\int \d x\, \bigl(R_{l+1}[u] + xu\bigr)\ .}
\par
In the following sections, we shall combine these properties with
a direct analysis of the differential equations
satisfied by the partition functions of the $d<1$ models to determine
the large order behaviour of the topological expansion of their solutions.
Previous work  \ref\rGZaplob{P. Ginsparg and J. Zinn-Justin, {\it Phys. Lett.}
B255 (1991) 189 and contribution to {\it Random surfaces and
quantum gravity\/}, proceedings of the 1990 Carg\`ese workshop, edited by
O. Alvarez, E. Marinari, and P. Windey, NATO ASI Series B262.} has allowed to
determine
that the topological expansion of the specific heat had the general property
of behaving like $a^k\Gamma(2k-1/2)$ for $k$, the order in the topological
expansion, large. The constant $a$ was determined as the solution of an
explicit algebraic equation for the one-matrix model ($q=2)$ and in two
examples the critical  and tricritical Ising model ($(3,4)$ and $(4,5)$
models). The importance of an explicit determination of $a$ relies on the
following property: If $a$ is real and positive the perturbative expansion
is not Borel summable and does not determine a unique function. Moreover there
are good reasons to expect the corresponding model to be actually unstable.
Such a result was obtained for half of the one-matrix models (this includes
pure gravity), and is expected for all unitary $(q+1,q)$ models.
This latter property is derived here and the more general models
$p=(2m+1)q\pm1 $ are explicitly discussed\foot{For a recent treatment of some
standard features of divergent series, Borel summability, and summation
methods, with physical applications, see, for example, pp.~840--842 of
\ref\rZJ{J. Zinn-Justin, {\it Quantum Field Theory and Critical Phenomena},
Oxford Univ. Press (1989).}.}.
\newsec{Large order behaviour of pure gravity}
We first recall the derivation of the large behaviour of pure gravity,
because it illustrates several features of the general analysis.
\par
For pure gravity, the differential equation satisfied by $u(x)$ is
\eqn\eai{u^2(x)-{1\over6}u''(x)=x\, .}
If $u(x)$ has an asymptotic expansion for $x$ large, it satisfies
$u(x)=\pm \sqrt{x}+O\left(x^{-2}\right)$.
The solution that corresponds to pure gravity has a $x$ large expansion of
the form
\eqn\eaii{u(x)=x^{1/2}\Bigl(1-\sum_{k=1}u_k\,x^{-5k/2}\Bigr)\ ,}
where the $u_k$ are all positive.\par
To determine the large order behaviour of the expansion we first
analyze the stability properties of the solution for $x$ large.
Let us set $u(x)\mapsto u(x)[1+\epsilon(x)]$ in eq.~\eai\ and write
the equation obtained by expressing that the term linear in $\epsilon$
vanishes:
\eqn\eepspg{\left(12u-{u'' \over u}\right)\epsilon-2{u'\over
u}\epsilon'-\epsilon''=0\,.}
One verifies that at leading order for $x$ large only the
leading order in eq.~\eaii\ is needed and  $u''/u$ is negligible.
Eq.~\eepspg\ can then easily be solved by the WKB method.
We set
$$\epsilon'/\epsilon=r\sqrt{u}+bu'/u+O\left(u'{}^2/u^{5/2}\right),$$
and find $r=2\sqrt{3}$ and $b=-5/4$. Replacing $u$ by its asymptotic form
$u\sim x^{1/2}$ and integrating we obtain:
\eqn\ews{\epsilon(x) \ \propto\
x^{-5/8}\,\ee{-{8\sqrt{3}\over5}x^{5/4}} .}
To leading order, the function $\epsilon$ is also proportional to the
difference between any Borel sum of the series and the exact non-perturbative
solution of the differential equation (up to even smaller exponential
corrections corresponding to multi-instanton like effects).
In terms of the expansion parameter (string loop coupling)
$\kappa^2=x^{-5/2}$, $\epsilon$ reads
\eqn\ewsp{\epsilon\bigl(x(\kappa)\bigr)
\ \propto\ \kappa^{1/2}\,\ee{-{8\over5}(\sqrt{3}/\kappa)}\ .}
The above solution is valid for $x$ large, i.e.\ $\kappa$ small. The large
order behaviour in \eaii\ is then given by
\eqn\elop{u_k
\mathop{\propto}_{k\to \infty}\ \int_0 {\d\kappa\over\kappa^{2k+1}}
\,\epsilon(\kappa)\ \propto\ \left({5\over8\sqrt3}\right)^{2k}
\Gamma(2k-\half)\ .}
(The constant of proportionality in the above
cannot be determined by this method.) The asymptotic $\Gamma(2k-\half)$
behaviour is a slight refinement of the $(2k)!$ behaviour determined in
\nref\rBIZ{D. Bessis, C. Itzykson, and J.-B. Zuber,
{\it Adv. Appl. Math.} 1 (1980) 109.}\refs{\rDS,\rGM,\rBIZ}.

The reality of $r^2$ has implied that all terms at large order have the
same sign. This induces a singularity on the real positive axis in the Borel
plane, obstruction to Borel summability.

In \ref\rFDii{F. David, {\it Nucl. Phys.} B348 (1991) 507.},
it is confirmed that the exponential in \ews\
coincides with the action for a single eigenvalue climbing to the top of
the barrier in the matrix model potential, allowing us to interpret the
exponential piece of the solution to \eai\ as an instanton effect.

\newsec{The general string equations}
\vskip-33pt
\subsec{The general one-matrix problem}
We now consider the string equation \ede, $R_l[u]\propto x$.
Substituting as before $u(x) \mapsto u(x)(1+\epsilon(x))$ we get a linear
equation for $\epsilon$. At leading order for $x$ large we expect the equation
to be again solved by the WKB ansatz $\epsilon'/\epsilon= r u^{1/2}$.
It is then easy to verify that to obtain the leading large order behaviour of
perturbation theory, it is only necessary to know the terms in $R_l[u]$
that contain at most one derivative of $u$ factor. The next leading
contribution is given by terms such as $u^{j-2}u^{(2l-2j-1)}u'$, i.e.\ with a
single factor of $u'$ as well. \par
At leading order only the terms in which the derivatives act on $\epsilon$
are relevant and thus $\epsilon$ satisfies an equation of the form
\eqn\elwkb{0=\sum^{l}_{j=1} A_{lj}\,u^{j-1}\epsilon^{(2l-2j)}\ .}
The WKB ansatz leads to an $(l-1)^{\rm st}$ order equation for the constant
$r^2$
\eqn\esa{0=A_l(r)\equiv\sum^{l}_{j=1} A_{lj}\,r^{2l-2j}\, .}
{}From the properties of the $R_l$'s one can derive an explicit expression for
the polynomials $A_l(r)$
 \eqn\eodra{A_l(r)\propto {1\over r}
\left(r^2-8\right)^{l-1/2}_+={\Gamma(l+1/2)\over
\Gamma(l)\Gamma(1/2)}\int_0^1 {\d
s\over\sqrt{s}}\bigl(r^2(1-s)-8\bigr)^{l-1}\, , }
where the subscript $+$ again means the polynomial part of the large $r$
expansion. The function $(z^2-1)^{l+1/2}_+$ is also proportional to
$C^{-l}_{2l+1}(z)$ where $C_{2l+1}^\nu$ is a Gegenbauer polynomial defined by
analytic continuation in $\nu$  \rGM. Note that the number of zeros is exactly
the same as the number of operators in a $(p=2l-1,2)$ minimal conformal model
\ref\rKPZ{V. G. Knizhnik, A. M. Polyakov, and A. B. Zamolodchikov,
{\it Mod. Phys. Lett.} A3 (1988) 819\semi
F. David, {\it Mod. Phys. Lett.} A3 (1988) 1651\semi
J. Distler and H. Kawai, {\it Nucl. Phys.} B321 (1989) 509.}.
This is a property we shall meet again in the general case. Actually in the
one-matrix case there is a natural explanation for it. The steepest descent
analysis shows that the number of different instantons is related to the
degree of the minimal potential corresponding to a critical point.
This degree in turn is also related to the number of relevant perturbations
\foot{We thank F.~David for this remark.}. \par
For $l$ even, eq.~\esa\ is an odd--order equation that will have at least one
real solution for $r^2$, positive as is obvious from the integral
representation \eodra. The series therefore cannot be
Borel summable.

For $l$ odd, on the other hand, the equation \esa\ for $r^2$ has no real
solutions and therefore we expect the solution of the differential equation
to be determined by the perturbative expansion.

Actually there exists a direct correspondence between the property of Borel
summability and the existence of the original integral. It has been noted
\ref\rBMP{E. Br\'ezin, E. Marinari,
and G. Parisi, {\it Phys. Lett.} B242 (1990) 35.}
that according to whether
$l$ is odd or even, the original minimal matrix integral is well-defined or
not because the integrand goes to zero in the first case while in the
latter case it blows up for $M$ large. A direct calculation, using
steepest descent, of the instanton action \rGZaplob\ confirms that when the
potential is unbounded from below the instanton action is indeed real and the
series therefore non-Borel summable. \par
The subleading terms in $R_l[u]$ are immediately deduced from the leading
terms by noting that since $R_l[u]$ is derived from an action, eq.~\egdp,
the operator acting on $\epsilon$ is hermitian.
Therefore the operator $u^{j-1}\d^{2l-2j}$ should be replaced by the
symmetrized form $\half\{u^{j-1},\d^{2l-2j}\}$, correcting \elwkb\ to
\eqn\elwkbp{0=\sum^{l}_{j=1} A_{lj}\bigl(u^{j-1}\epsilon^{(2l-2j)}
+\half(2l-2j)(j+1)u^{j-2}\,u'\,\epsilon^{(2l-2j-1)}\bigr)\ .}
To characterize more precisely the large order behaviour, to next
order we set $\epsilon'/\epsilon = ru^{1/2} + b u'/u$,
from which it follows, to the same order, that
\eqn\ewkbno{{\epsilon^{(k)}\over\epsilon}
= r^k u^{k/2} + r^{k-1} u^{(k-3)/2}\,u'\,k\Bigl(b+\frac{1}{4}(k-1)\Bigr)\ .}
Substituting into \elwkbp, we find $b=-(2l+1)/4$, independent of $r$.
Then
\eqn\egeps{\epsilon(x) \ \propto\ x^{-(2l+1)/4l}\,\ee{-{2l\over2l+1} r
x^{(2l+1)/2l}}\ ,}
generalizing \ews.
In terms of the expansion parameter $\kappa=x^{-(2l+1)/2l}$, we find
\eqn\elobg{u_k\ \mathop{\propto}_{k \to \infty}\ \int_0
{\d\kappa\over \kappa^{2k+1}}\epsilon\bigl(x(\kappa)\bigr)
\ \propto\ \left({2l+1\over2lr}\right)^{2k}\Gamma(2k-\half)\, .}
The $\Gamma(2k-\half)$ factor in \elop\ is general, and is related to the
property that the original equations descended from an action principle.

\subsec{The general $(p,q)$ model}
In the case of the general $(p,q)$ model (eq.~\egenac) there
results a system of coupled linear differential equations for the
variations $\epsilon_{u_i}(x)$ associated with the functions
$u_i(x)$.  At leading order we set $\epsilon'_{u_i} /\epsilon_{u_i}=r u^{1/2}
$. We obtain, taking into account the leading relations between the
$u_i$, a linear system for each of the $\epsilon_{u_i}$'s multiplied by a
power of $u$ determined by the grading.  Imposing
again the vanishing of the determinant of the linear system gives an
equation for the coefficient $r$ (and to leading order all functions
$\epsilon_{u_i}$ are thus proportional up to a power of $u$ determined by the
grading).
\nref\rGZ{P. Ginsparg and J. Zinn-Justin, {\it Phys. Lett.} B240 (1990) 333.}%
\nref\rdOne{E. Br\'ezin, V. A. Kazakov, and Al. B. Zamolodchikov,
{\it Nucl. Phys.} B338 (1990) 673\semi
G. Parisi, {\it Phys. Lett.} B238 (1990) 209, 213;
{\it Europhys. Lett.} 11 (1990) 595\semi
D. J. Gross and N. Miljkovic, {\it Phys. Lett.} B238 (1990)
217.}

To determine more precisely the behaviour of $\epsilon_u=\epsilon_{u_1}$ we
have to consider subleading terms. As in the one-matrix case they can be
determined by a hermiticity argument.
Since the equations for $u_i$ derive from an action \egenac, the linear
equations for $\epsilon_{u_i}$ define a hermitian operator. This property
leads to a universal $\Gamma(k-\half)$ behaviour for all the $(p,q)$ models.
\par
We recall finally that the $(2k)!$ large order behaviour
is also the generic behaviour \rGZ\ for $d=1$ models \refs{\rGZ,\rdOne}.

\newsec{The $p=(2m+1)q\pm1 $ models in the spherical limit}
In the analysis of the large behaviour the knowledge of the solutions of the
string equation in the spherical limit was required. Actually we shall prove
in next section that the knowledge of the differential operators $P,Q$
in the same limit is sufficient. For $q=2$ and $q=3$ the form of the operator
$Q$ and $P$ is known. In the general case $q\ge 4$ the explicit functional
form of the operator $Q$ in the spherical limit depends on the $(p,q)$ models.
In particular in the spherical or semiclassical limit the operator $Q$ takes
the form
\eqn\eompq{Q(\d,u)=\sum_{i=0}q_i u^i(x)\d^{q-2i},}
but the coefficients $q_i$ in equation \eompq\ are in general $p$-dependent.
Note that, in this limit, the order between the operators $u(x)$ and $\d$ is
irrelevant. \par
However in \ref\rEyZJ{B. Eynard and J. Zinn-Justin, {\it Nucl. Phys.} B386
(1992) 558.} it has been shown, using the string actions \egenac\
\rGGPZ\  in the semiclassical limit, that when $p=(2m+1)q\pm1$, the
semiclassical form of the operator $Q$ is $m$-independent and $P$ and $Q$ can
be determined explicitly. This property can be recovered by a direct method.
If we also set:
\eqn\ePsmcl{P(\d,u)=\sum_{i=0}p_i u^i(x)\d^{p-2i},}
we obtain the semiclassical limit of the equation $[P,Q]=1$:
\eqn\ecomPQ{u' \left({\partial P\over \partial \d}{\partial Q\over \partial u}
-{\partial Q\over \partial \d}{\partial P\over \partial u}\right)=1\,.}
We now use the homogeneity property of $P,Q$:
$$P(\d,u)=u^{p/2}P\left(\d u^{-1/2},1\right) ,\qquad Q(\d,u)
=u^{q/2}Q\left(\d u^{-1/2},1\right) .$$
{}From now on we call $P(z)$, $Q(z)$ the two polynomials $P(z=\d u^{-1/2},1)$,
$Q(z=\d u^{-1/2},1)$. They thus satisfy the differential equation:
\eqn\ecomPQii{qP'(z)Q(z)-pP(z)Q'(z)=2pq\,,}
while as expected the equation for $u(x)$ yields $u^{(p+q-1)/2}\propto x$.
When one of the polynomials is known the other is obtained by integrating the
equation. In the special case $p=(2m+1)q\pm 1$
the polynomials $Q(z)$ are
Tchebychev's polynomials. Setting $z=2\cos\theta$ one finds that $Q(z)
= 2\cos q\theta$ and
$$P=2p Q^{p/q}(z)\int_0^z Q^{-1-p/q}(t)\d t\propto 2\sum_{l=0}^m  {p/q \choose
l}\cos(p-2lq)\theta\,,$$
satisfy the equation.
\newsec{Instantons: A more direct method}
\vskip-33pt
\subsec{Instantons in the one-matrix model revisited}
Before discussing the general unitary models let us return to the one-matrix
model for which the result is exactly known. From the analysis of the
corresponding non-linear differential equations we have learned that if we
call $\epsilon$ the variation of the specific heat $u(x)$ then it has
for $x$ large the asymptotic form:
\eqn\eepscom{\epsilon'/\epsilon\sim r \sqrt{u},}
where $r$ is constant which is determined by an algebraic equation. Since
the variation of $u$ can be neglected at leading order, we can rescale $\d$,
i.e.\ set $u$ to 1. Equation \eepscom\ can then be written as a commutation
relation
\eqn\eepscomii{\d \epsilon=\epsilon(\d+r)\ \Rightarrow\
f(\d)\epsilon=\epsilon f(\d+r).}
Then the operators $P,Q$ are simply
$$Q=\d^2-2\,,\qquad P\equiv P_{2l+1}(\d)=\left(d^2-2\right)^{l+1/2}_+.$$
The equation  for $\epsilon$ is obtained by expanding at first order in
$\epsilon$ the commutation relation $[P,Q]=1$. Setting:
$$\delta P=\{\epsilon,R(\d)\}\equiv\sum_{k=0} R_k \{\epsilon,
d^{2l-1-2k}\},$$
we find:
$$[\{\epsilon,R(\d)\},\d^2-2]+[P,-2\epsilon]=0\,.$$
Using the commutation relation \eepscomii\ to commute $\epsilon$ to the left
we find the equation:
$$-\left(2r\d+ r^2\right)\bigl(R(\d)+R(\d+r)\bigr)-2
\bigl(P_{2l+1}(\d+r)-P_{2l+1}(\d)\bigr)=0\,.$$
The first term vanishes for $\d=-r/2$, which must thus be a zero of the second
term. Taking into account the parity of $P_{2l+1}$ we obtain
\eqn\eonemat{P_{2l+1}(r/2)=0\,,}
in agreement with the direct calculation. The polynomial $R(\d)$ is then
determined by division.
\subsec{General $(p,q)$ problem}
In the general $(p,q)$ case, in the same classical limit, and after the same
rescaling we have:
$$Q=Q(\d),\quad P=P(\d)=Q^{p/q}_+(\d),\quad \delta Q=\{S(\d),\epsilon\},\quad
\delta P=\{R(\d),\epsilon\},$$
where $P$, $Q$ are polynomials of degrees $p$, $q$ respectively,
and $R$, $S$ of degrees $p-2$, $q-2$ and same parity as $P$, $Q$.\par
The equation for $\epsilon$ then leads to
$$\eqalignno{[P,\delta Q]+[\delta P,Q]&=0 \cr\Leftrightarrow \
\bigl(P(\d+r)-P(\d)\bigr) \bigl(S(\d)+S(\d +r)\bigr) & \cr -
\bigl(Q(\d+r)-Q(\d)\bigr) \bigl(R(\d)+R(\d +r)\bigr)&=0\ .\cr}$$
The polynomials $P(\d+r)-P(\d)$ has a degree $p-1$ in $\d$, while $R$ has only
a degree $p-2$. An equivalent property is true for $Q,S$. The polynomials
$P(\d+r)-P(\d)$ and $Q(\d+r)-Q(\d)$ must thus have at least one common root.
Note that the first polynomial has $p-1$ roots and the second $q-1$. Moreover
this roots are symmetric in the exchange $\d\mapsto -r-\d$. Therefore
expressing the existence of a common root leads to $(p-1)(q-1)$ values
of $r$, up to the symmetry. Note that the number of zeros is again exactly
the same as the number of relevant operators in a $(p,q)$ minimal conformal
model \rKPZ\ of gravitationally dressed weights
\eqn\ekpz{d_{m,n}={p+q-|pn-qm|\over p+q-1}\ ,}
with $1\le n\le q-1$, $1\le m\le p-1$ with the symmetry
$d_{m,n}=d_{q-n,p-m}$. The explanation of this relation is probably again
that the number of different instanton actions is related to the degree
of the minimal potential needed to generate a critical point in the
multi-matrix model, and thus to the number of different relevant operators.
Also we note that we are studying a general deformation of a critical
solution and therefore the appearance in some form of the relevant operators
should be expected.\par
This condition determines the possible  values of $r$ when the polynomials
$P$ and $Q$, i.e.\ the differential operators are known in the classical
limit. Examples are provided by the models $p=(2m+1)q\pm 1$ where integral
representations for these polynomials have been found. The simplest examples
are provided by the $q+1,q$ models, i.e.\ the unitary models which we
examine below. \par
Let us finally verify that we can then indeed find the polynomials
$R,S$. We call $\alpha$ the common root and assume first that $\alpha\neq
-r/2$. Then the parity properties imply that $-r-\alpha$ is also a common
root. Setting then
$$\eqalign{\bigl(P(\d+r)-P(\d)\bigr)&=(\d-\alpha)(\d+r+\alpha)\tilde
P(\d), \cr
\bigl(Q(\d+r)-Q(\d)\bigr)&=(\d-\alpha)(\d+r+\alpha)\tilde
Q(\d), \cr}$$
we find that $R$ and $S$ are solutions of:
$$R(\d)+R(\d +r)=(\d+r/2)\tilde P(\d),\quad S(\d)+S(\d +r)=(\d+r/2)\tilde
Q(\d).$$
Note that these equations satisfy both the degree and parity requirement.
\par
If $\alpha=-r/2$ the situation is even simpler
$$\eqalign{ R(\d)+R(\d +r)&=\bigl(P(\d+r)-P(\d)\bigr)/(\d+r/2) \cr
S(\d)+S(\d +r)&=\bigl(Q(\d+r)-Q(\d)\bigr)/(\d+r/2). \cr}$$
\subsec{The unitary models}
We have shown that the differential operators $P,Q$ may be written in the
classical limit as:
$$P= 2 T_p(\d/2)\,, \qquad Q=2 T_q(\d/2), $$
where $T_p$ is the $p$-th Tchebychev's polynomial:
$$T_p(\cos\varphi)=\cos{p\varphi}.$$
As explained above, taking into account the degrees of the polynomials $R$
and $S$, we conclude that the polynomials $T_q((r+\d)/2)-T_q(\d/2)$ and
$T_p((r+\d)/2)-T_p(\d/2)$  must have a common root $\alpha=2\cos{\varphi_0}$.
Let also set $\alpha+r=2\cos{\psi_0}$.
We have
$$\cos{p\psi_0}=\cos{p\varphi_0} \quad {\rm and} \quad
\cos{q\psi_0}=\cos{q\varphi_0}.$$
The solution is :
$$\psi_0 = \pm \varphi_0 + {2m\pi\over p} = \mp \varphi_0 + {2n\pi\over q}$$
Since $r=2\cos{\psi_0}-2\cos{\varphi_0}$, excluding the solutions $r=0$
which is not acceptable, we have the different solutions:
$$ r=\pm 4\sin{m\pi/p}\sin{n\pi/q},\quad 0< 2m  \le p\,,\quad 0< 2n \le q \,
.$$
It is easy to verify that these results agree with the explicit solutions of
the $(2,3)$, $(4,3)$ and $(4,5)$ models. They show that, as expected, all
unitary models lead to non-Borel summable topological expansions because all
terms of the series have the same sign. These models thus suffer from the same
disease as the pure gravity model. Note finally that indeed the number of
different values of $r$ is the same as the number of operators in the
minimal $(p,q)$ conformal model.
\subsec{The $p=(2m+1)q\pm 1$ models}
For $m\neq 0$ $r$ is solution of more complicated algebraic equations.
In the notation of previous subsection we still have:
$$\psi_0=\pm \varphi_0+{2n \pi \over q}.$$
Let us set
$$\alpha=\ud (\psi_0+\varphi_0),\qquad \beta=\ud (\psi_0-\varphi_0),$$
then, making a choice of signs
$$\beta={n \pi \over q},\qquad r=4\sin\alpha\sin \beta=4\sin\alpha\sin(n\pi/q)
,$$
where
$$\sum_{l=0}^m {p/q\choose l}\sin[(p-2ql)\alpha]\sin[(p-2ql)\beta]=0\,.$$
We note that $\sin[(p-2ql)\beta]=\sin(n\pi p/q)$ which can be factorized.
We thus find an equation for $\alpha$:
$$A(\alpha)\equiv \sum_{l=0}^m {p/q\choose l}\sin[(p-2ql)\alpha]=0\,.$$
This function satisfies the differential equation
$$p A(\alpha)(\cos q\alpha)'-q A'(\alpha)\cos q\alpha= K(p,q)\cos \alpha\,,$$
where $K$ is a constant. This equation implies that $A(\pi/2q)A(3\pi/2q)\le 0$
and thus $A(\alpha)$ vanishes at least once in the interval $(0,\pi)$. We
conclude that  for all these models the topological expansion is not Borel
summable.
\listrefs
\bye